\documentclass[conference]{IEEEtran}
\usepackage{etex}
\makeatletter
\def\ps@headings{%
\def\@oddhead{\mbox{}\scriptsize\rightmark \hfil \thepage}%
\def\@evenhead{\scriptsize\thepage \hfil \leftmark\mbox{}}%
\def\@oddfoot{}%
\def\@evenfoot{}}
\makeatother
\ifx\pdfoutput\undefined
\usepackage{graphicx}
\else
\usepackage{slantsc}
\usepackage[pdftex]{graphicx}
\usepackage{epstopdf}
\fi 
\usepackage{algorithm}
\usepackage{varwidth} 
\usepackage{algpseudocode}
\usepackage[english]{babel}
\usepackage{url}
\usepackage{verbatim}
\usepackage{caption}
\usepackage{subfigure}
\usepackage{blindtext}
\usepackage{amssymb}
\usepackage{lipsum,multicol}
\usepackage{fmtcount}
\usepackage[numbers, square, comma, sort&compress]{natbib}
\usepackage[fleqn]{amsmath} 
\usepackage{tikz} 
\usepackage{colortbl}
\usepackage{array,booktabs,arydshln,xcolor} 
\usepackage[bottom]{footmisc}
\usepackage{authblk}
\usetikzlibrary{automata,positioning} 
\title{QoE-Centric Localized Mobility Management for Future Mobile Networks}

\begin{document}
\algnewcommand\algorithmicswitch{\textbf{switch}}
\algnewcommand\algorithmiccase{\textbf{case}}
\algnewcommand\algorithmicassert{\texttt{assert}}
\algnewcommand\Assert[1]{\State \algorithmicassert(#1)}%
\algdef{SE}[SWITCH]{Switch}{EndSwitch}[1]{\algorithmicswitch\ #1\ \algorithmicdo}{\algorithmicend\ \algorithmicswitch}%
\algdef{SE}[CASE]{Case}{EndCase}[1]{\algorithmiccase\ #1}{\algorithmicend\ \algorithmiccase}%
\algtext*{EndSwitch}%
\algtext*{EndCase}%
\author[ ]{Esmat Mirzamany}
\author[1]{Vasilis Friderikos}
\affil[1]{Centre for Telecommunications Research\\ King's College London\\}
\affil[ ]{\textit {e\_{}mirzamany@yahoo.com, vasilis.friderikos@kcl.ac.uk}}
\renewcommand\Authands{ and }
\maketitle
%
\begin{abstract}
Mobility support in future  networks will be predominately based on  micro mobility protocols. Current proposed schemes such as Hierarchical Mobile IPv6 (HMIPv6) and more importantly Proxy Mobile IPv6 (PMIPv6) provide localized mobility support by electing a node within the network (topologically close to the Access Routers (AR)) to act as mobility anchor . Such schemes can significantly improve handover latency, as well as the  end-to-end signalling overhead, but might entail scalability issues, which in some instances, do not fit adequately with the current explosion of mobile Internet traffic, and the evolutionary trend towards flat network architectures. The notion of using Distributed Mobility Management (DMM) allows for decentralization by anchoring nodes at their AR. The idea is that for sessions with duration less than the cell residence time efficient mobility can be supported. However, DMM might be highly suboptimal in instances where nodes perform multiple handovers during the session lifetime. Hence, one approach cannot be effective for all types of applications. In this paper a hybrid mobility management solution, integrated with a new routing scheme, is proposed. The scheme selects the most suitable mobility approach, centralized or distributed, by taking into account the flows' requirement, in terms of Quality of Experience (QoE), and new routing constraints. The results of optimization problem show that the proposed approach can achieve efficient resource utilisation, ease network congestion, and lead to a significant improvement in the \looseness=-1 QoE.
\end{abstract}
\section{INTRODUCTION}
Two crucially important supported network functionalities for emerging heterogeneous cellular/wireless networks are localized IP based mobility and the support of Quality of Experience (QoE). The former will in essence allow seamless mobility of users across various wireless access technologies, while the latter would allow a user-centric network management and resource allocation. 
Defined by ITU-T, QoE is the overall acceptability of an application or service, as perceived subjectively by the end-user \cite{QoEitu-trecommendation}. While the Quality of Service (QoS) has been extensively studied, concentrating on the actual QoE is a more recent trend that takes user perception of the delivered service into account \citep{Yen2013,Imran2013,Taboada2013}. Hence, the relationship between QoS parameters and their effect on the QoE has been    intensively studied recently \citep{Fiedler2010,Chen2012}. 
At the same time, supporting IP level mobility is gaining significant momentum as the heterogeneity of wireless connectivity is increasing. Both host- and network-assisted localized mobility support, mainly referred to as Centralised Mobility Management (CMM), are based on the idea of deploying mobility anchoring at nodes within the scope domain. Routing all aggregate traffic through a mobility anchor imposes unbalanced use of available network resources (in terms of the use of available routing paths), and vulnerability to single points of failure. To alleviate some of these concerns, a Distributed Mobility Management (DMM) scheme has been recently proposed \citep{Chan2014}. In DMM, mobility anchoring is taking place at the AR instead  of a node within the core/access network. The aim of DMM is to capitalize instances where the session duration is estimated to be less than the cell residence time; meaning that only one handover, if any, is envisioned during the session. Hence, DMM deployment provides an optimal initial anchor assignment but for users with increased mobility and frequent changes of the point of attachment the performance can deteriorate \citep{Liu2014}.  Therefore, while the CMM efficient handover management makes it competent for applications with tight delay constraints, the opportunity of having a more elastic path selection offered by the DMM can make it more suitable for delay-tolerant \looseness=-1 applications. 

Based on the above insights, we propose a hybrid mobility management scheme to select either anchoring within the network or at the AR, by taking into account the requirements and characteristics of the traffic. To that end, a mobility anchor located within a domain is assigned for real-time traffic, while the non real-time traffic is anchored to the associated ARs. Hereafter, a new mobility aware routing algorithm is proposed which routes  traffic through the longest acceptable paths, making the shortest ones available to the delay-sensitive traffic. Therefore, by making use of all available paths, the maximum congestion in the network can be reduced, especially at the anchor which is within the network.
The proposed scheme is in essence inspired by the Multi-Topology Routing (MT-Routing) concept \citep{rfc4915}, which provides routers with multiple logical views of the network, each one with an independent set of link weights. The concept is that firstly edge-disjoint logical topologies of the network are discovered by proposed heuristic algorithms which are detailed in the sequel sections. Thereafter, based on a defined optimization problem, a logical topology which encapsulates the \textit{longest acceptable path} is used to route the flows. 
In order to ensure a minimum acceptable service level for low-priority traffic, the worst-case performance is bounded, ensuring that a longest path delay does not exceed a certain maximum \looseness=-1 threshold value.

The remainder of the  paper is organised as follows: The system model and problem definition are introduced in Section II. The proposed algorithms for building disjoint routing topologies and selecting the most efficient ones are detailed in Section III. The performance analysis of the proposed scheme is discussed in Section IV. Finally, the conclusions in section V summarize the main findings of this paper.
\section{SYSTEM MODEL}
A mobile core/access network is represented by a weighted directed graph $G=(V,E,c,d)$, where $E$ represents the set of links and $V$ is the set of nodes, including relays, eNBs and routers. The delay and capacity of a link connecting node $i$ to node $j$ are represented by $d(i,j)$ and $c(i,j)$, \looseness=-1 respectively. If we denote  a path between an origin $s$ and a destination $t$  by $p$, the delay across the path, which is an additive metric, is:
\begin{equation}
d(p)= \sum_{(i,j)\in p} d(i,j)
\label{del-bw}
\end{equation}
The traffic matrices are represented by $R=\{{}r_{st}\mid s,t \in V\}{}$, where $r_{st}$ reflects the traffic demand between a given origin-destination (O-D) pair. For each pair node, a link-based routing $x$ is defined by a set of variables $x=\{{} x_{ab}(i,j)\mid a,b,i,j\in V\}{}$, where $x_{ab}(i,j)$ is a fraction of traffic demand between a pair $a\rightarrow b$, that goes through the link $(i,j)$. 
%
We further assume that there are $K$ disjoint logical topologies each one containing a path $p^k$ between the O-D pair indexed by $(s,t)$. Each $p^k: k=1,2,...,K$ is associated with a delay $d^k(p)$, obtained from equation \ref{del-bw}. A traffic demand $r_{st}$ with the delay constraint $D^\tau$ can be routed through the network if there is a path in one of the topologies whose delay is less than or equal to the flow's delay constraint. 
With these insights, the proposed scheme can be formulated as a constrained optimization problem with the objective function of maximizing the minimum residual capacity in the network, leading to an increase of the available resources in the bottleneck link, which in essence reflects the mobility anchor. The optimization problem with its objective and constraints can be formulated as follows:
\begin{equation}
\small
\begin{split}
&Maximize ~ Min_{(i,j)\in E}\big(C(i,j) - f(i,j)\big) \\
&\mbox{~subject to:~}\\
&\sum_{j:(i,j)\in E}\sum_{k}x_{st}^k(i,j)-\sum_{j:(j,i)\in E}\sum_{k}x_{st}^k(j,i)=\left\{\setlength\arraycolsep{1pt}\begin{array}{rr}
-1&\mbox{$i=t$}\\
0&\mbox{ $i\neq s,t$}\\
1&\mbox{$ i=s$}
\end{array} \right. \\
&\forall s,t\in V: \sum_{k\in K} x^k_{st}=1 \quad \\ 
&\forall (i,j)\in E, r\in R,and \quad s,t\in V: \sum_{st} x^k_{st}(i,j)r_{st}=f_{i,j}^k \quad \\
&\forall(i,j)\in E: 0 \leq x_{st}^k(i,j)\leq 1 \\
&\forall(i,j)\in E: 0 \leq f_{i,j}\leq C_{i,j} & \\
\label{Slave-Master}
\end{split}
\end{equation}
\begin{equation*}
\small
\begin{split} 
%
%
%
&\forall s,t\in V, r\in R, k\in K: \mbox{~Maximize~} \sum_{k\in K}O(r_{st}^\tau ,k) \\ 
&\quad \mbox{~where~~} O(r_{st}^\tau,k)=\sum_{(i,j)\in p^k}\frac{d(i,j)}{D^\tau} \leq \Gamma
\label{Slave-Master}
\end{split}
\end{equation*}
%
The first constraint satisfies the flow conservation. The second constraint guarantees that the sum of all of the traffic's fractions routed through different topologies is conserved, i.e., is equal to one. The third constraint depicts that the total load of a topology $k$, passing through link $(i,j)$, is equal to the sum of all its fractional traffic (between all the O-D pairs) routed through this link. 
The last constraint in equation \ref{Slave-Master} ensures that flows are routed through the longest acceptable path, keeping the shortest paths available for the most time-critical flows. Assuming furthermore that the link delay and the Service Level Agreement (SLA)-based delay constraint $D^\tau$ are non-negative, $\Gamma$ is bound to fall in the $(0,1]$ interval. 
Since formulation \ref{Slave-Master} does not comply with the standard linear programming setup, the last constraint can be firstly solved using the following \textit{slave} linear optimization \looseness=-1 problem. 
\begin{equation}
\begin{split}
&\mbox{~Maximize~} \quad\sum_{k\in K} \sum_{(i,j)\in p^k}\frac{d(i,j)}{D^\tau}\\
&\mbox{~subject to:~}\\
&\quad \forall k\in K: \sum_{(i,j)\in p^k} d(i,j) \leq D^\tau \quad \\
&\quad \forall (i,j)\in E: d(i,j)> 0, D^\tau > 0
\label{Slave}
\end{split}
\end{equation}
In order to ensure that the low-priority traffic can still get an acceptable performance, the proposed algorithm, which is discussed later in the next section, guarantees that the delay of the longest path always remains lower than the acceptable upper bound delay.
In essence, delays on a the link $d(i,j)$ can be decomposed into the propagation, transmission and queuing delays. Assuming that all the nodes are connected through identical high-capacity links, the transmission delay can be considered as fixed and smaller compared to the propagation delay, for all the en-route nodes. Moreover, the maximum queueing delay for a buffer size $B$ is equal to $\frac{B}{C}$, where $C$ is the link capacity. The length of $B$ is defined by a widely used rule-of-thumb, which states that each link needs a buffer of size $B=RTT\times C$, where $RTT$ is the average round-trip time of a flow passing through the link \citep{Appenzeller2004}. 
Nevertheless, recent studies show that buffers could be significantly smaller, perhaps as small as a few $\mu s$. Their argument is based on the fact that large buffers increase queuing delay at a congested link, which results in an increase in the round trip time. Therefore, today's Internet with dominant TCP based real-time traffic is better off with lower latency than lower drop probability, since applications can protect themselves against packet
drops but lost time can never be recaptured \cite{Wischik2005}. With this insight, we consider the propagation delay as a dominant component in $d(i,j)$ throughout our study. 
\begin{figure*}[htp]
\centering
\subfigure[Graph G where each link is represented by ($c,d$)]{
\includegraphics[scale=.50]{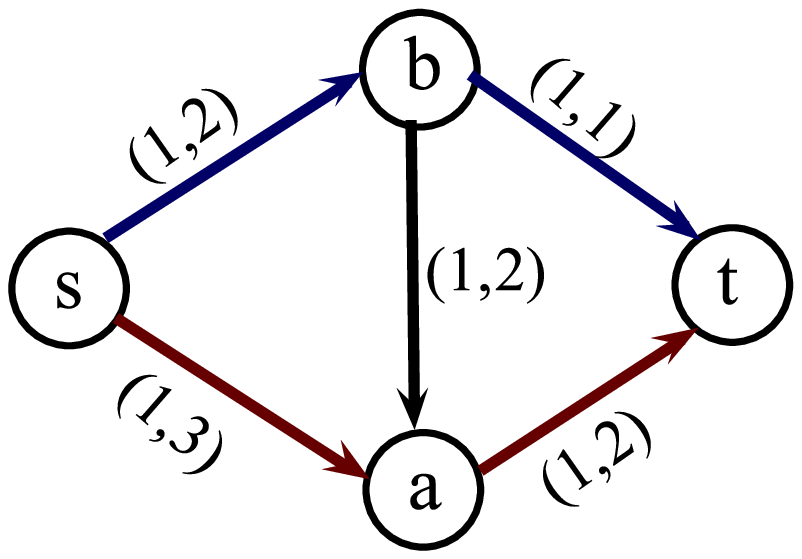}
\label{fig:SampleG}
} \hspace{2cm}
\subfigure[Graph $G^D$ with D=6]{
\includegraphics[scale=.35]{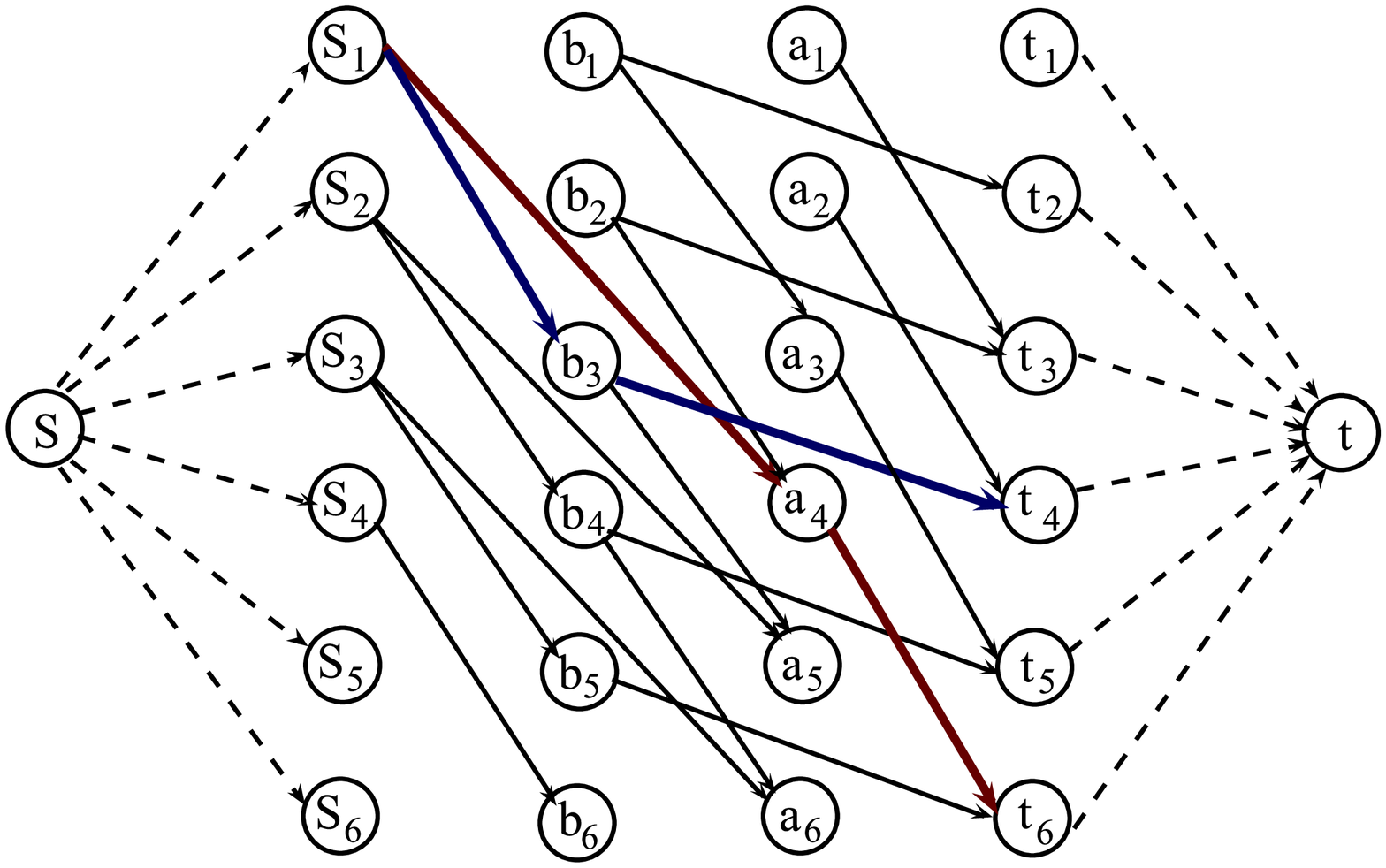}
\label{fig:MakingGD}
}
\caption[Making Graph]{Constructing a layered graph $G^D$ from a graph $G$}\label{fig:MakingG}
\end{figure*}
\section{FINDING EDGE-DISJOINT LOGICAL TOPOLOGIES}
In this section heuristic algorithms, to find a set of disjoint paths with the maximum delay of $D$ and select the best candidate in the set, are proposed. The first algorithm is based on a graph transformation technique used by \citep{Ahuja1993}.
Let an instance of the network be given by the graph $G=(V,E,c,d)$ with the maximum delay constraint $D>0$ and O-D pair $(s,t)$, while $s$ is a current node and $t$ can be any given node in the network. Both $D$ and the link delay are assumed to be integers. 
The algorithm aims to construct a \textit{layered graph} $G^D=(V^D,E^D)$ from $G$ in the following way:
\begin{itemize}
\item Make $D$ copies $u_{1},u_{2},...,u_{D}$ of each node $u\in V$. Node $u_k$ in the $G^D$ represents node $u$ of the original network at time $k$.
\item Include link $(i_k, j_l)$ of capacity $c(i,j)$ in $G^D$ whenever link $(i,j)\in E$ and $l-k=d(i,j)$. The link $(i_k,j_l)$ in $G^D$ represents the potential movement of a commodity from node $i$ to node $j$ in $d(i,j)$ time.
\item Reduce the multiple-source, multiple-sink problem in $G^D$ to the single-source, single-sink problem by introducing a super-source $S$ and super-sink $t$ with the set of links, each with the bandwidth $\infty$.
\end{itemize}
It is straightforward to see that any path between origin $s_1$ and destination $t$ in $G^D$ has a corresponding path $p\in G$ such that $d(p)<D$. The constructed layered graph $G^D$ contains all the disjoint paths, with the delay less then $D$, between $S$ and all possible destinations in the network. By selecting any given node $u$ as a destination, the set of $u_{1},u_{2},...,u_{D}$ can be reduced to a super-sink, with a set of links each with infinite bandwidth capacity. Therefore, by moving from $s_1$ to the super-sink, all possible disjoint paths can be discovered. An example of the graph $G$ and its transformation to the graph $G^D$ with $D=6$ is given in figure \ref{fig:MakingG}. 
After constructing $G^D$, algorithm \ref{MDisjoint} is used to find all existing disjoint logical topologies for a given network. For each link coming out of $s_1$, the shortest path between $s_1$ and a given destination $t$ is calculated, and the corresponding links are added to the $G_{[K]}$ (lines 6-7). The process continues until all of the other nodes are selected as a destination and their disjoint paths are discovered (loop in line 5). After adding the corresponding links to the $G_{[K]}$, these links are deleted from the graph $G^D$ (line 9), guaranteeing that they will not be used in other topologies. The same process is repeated for all of the outgoing links from \looseness=-1 $s_1$. 
\begin{algorithm}[hbp]
\caption{Finding K Disjoint Logical Topologies $G_{[K]}$ }\label{MDisjoint}
\begin{algorithmic}[1]
\State Construct $G^D$ from $G$ using the proposed graph transformation technique
\State $K=0$
\While{$outdegree(s_1)\neq 0$} 
\State $K=K+1$
\While{all nodes being selected as a destination t}
\State find the shortest path $p_K$ between O-D pair $(s_1,t)$
\State Put $\forall (u,v)\in p_k$ in $G_{[K]}$
\EndWhile\label{allnodesDes}
\State Remove $\forall (u,v)\in p_K$ from $G^D$ 
\State $outdegree(s_1)=outdegree(s_1)-1$
\EndWhile\label{euclidendwhile}
\State \textbf{return} $G_{[k]}, k=1,...,K$ as the set of disjoint logical topologies 
\end{algorithmic} 
\end{algorithm}
%
%
We note that fully disjoint topologies reduce the size of routing tables associated with each topology, while at the same time such operation increases  network resiliency. 
Note that if any link is left out, it is included in the topology of which its parent link belongs, as long as the maximum delay constraint of the path is not violated. Adding these links can increase the number of possible routes in some part of each topology, resulting in a better link utilisation. 
With respect of the complexity, the procedure can be deemed as efficient since in essence only a limited number of shortest paths are created.

\subsubsection*{\textbf{Finding Efficient Logical Topologies}}
After finding all the edge-disjoint topologies in the network, the possible candidates need to be chosen for a given flow, in a way that each one complies with the flow's requirement. Algorithm \ref{FindingBestMT} describes how the best set of logical topologies would be chosen by taking into the account the flow's traffic class $\tau$ and its SLA based delay constraint $D^\tau$. 
\begin{algorithm}
\caption{Finding Best Topology for Flow $(r^\tau_{s_1t},D^\tau)$ }\label{FindingBestMT}
\begin{algorithmic}[1]
\For { each $G_{[k]}$ }
\If {$\frac{d(p_{\mathtt{k}})}{D^\tau} \leq 1$}
\State Add $G_{[k]}$ to the set of feasible solutions $\chi_r$
\State $\chi_r = \chi_r+1$
\EndIf
\EndFor
\Switch{$\chi_r$}
\Case{$\|\chi_r\|=0$}
\State No feasible path is found to comply with SLA
\EndCase
\Case{$\|\chi_r\|=1$}
\State \begin{varwidth}[t]{\linewidth}
Send the flow $(r^\tau_{s_1t},D^\tau)$ via only possible topology
\end{varwidth} 
\EndCase
\Case{$\|\chi_r\|>1$}
\State \begin{varwidth}[t]{\linewidth}
Select a topology with the highest $\frac{d(p_{\mathtt{k}})}{D^\tau}$ value
\end{varwidth}
\EndCase
\EndSwitch
\State \textbf{end switch} 
\end{algorithmic} 
\end{algorithm}

Based on the proposed algorithm, for each logical topology if the delay of the path between two end points is less than the flow's delay constraint, the topology is added as one possible solution (line 2). At the end, if there is more than one candidate, the path with the highest accepted delay is  selected.
It is important to emphasis that the main objective of the proposed scheme (as detailed in ~\ref{Slave-Master}) is to ease the congestion in the network by shifting the traffic from highly-loaded links to the lightly utilized ones. This can lead to a more balanced link utilization in the network. Also provision of adequate resources for the different services in the network should be implemented when for example the longest acceptable paths of delay elastic traffic overlaps with the shortest path assigned to delay sensitive flows. 
\section{PERFORMANCE EVALUATIONS}
\begin{figure}[tp]
\includegraphics[scale=.22]{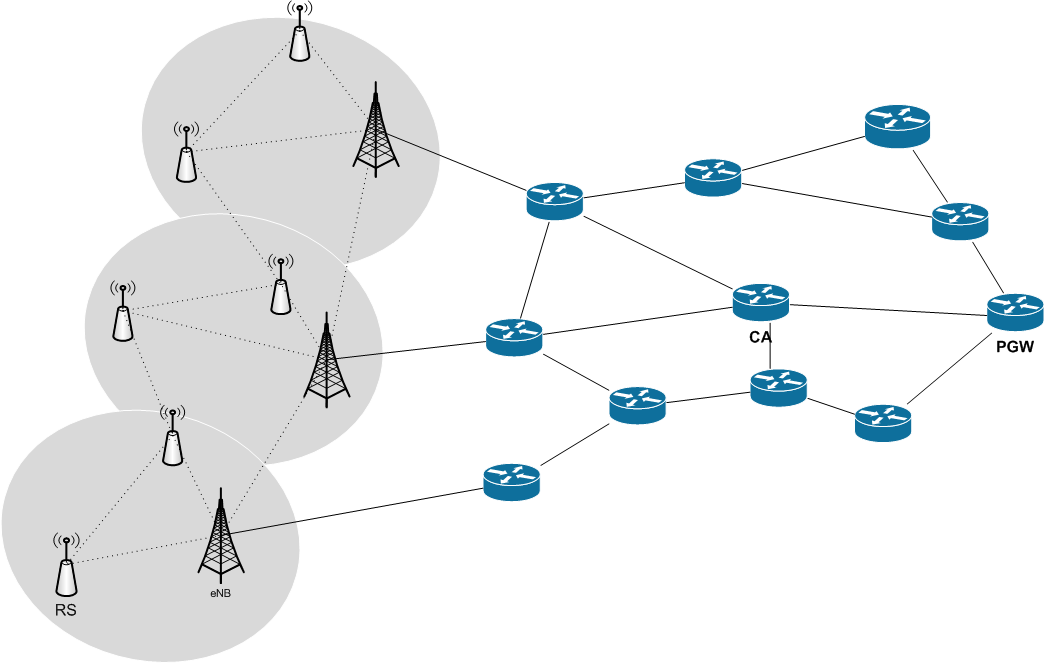}
\caption{Network Topology}\label{fig:Topology1}
\end{figure}

In this section we study the performance of the proposed scheme as compared to other routing algorithms. The network topology, depicted in figure~\ref{fig:Topology1}, consists of three typical cells in an LTE network, each having one 15 MHz 2x2 eNB and two fixed layer-3 relays located at the cell edge. For a given configuration, the backhaul wired links capacity between eNBs and PDN Gateway (PGW) is set to 40Mbps \citep {Cambridge2011}, while the wireless connection between an eNB-eNB and a relay-eNB is set to 20Mbps. All the connections have a delay of 1ms. The router CR acts as a central anchor for real-time traffic, while the ARs are used as a distributed anchor for non real-time traffic.
The center cell is assumed to be the most congested one, serving 450 simultaneous users, while each of the other cells is assumed to serve 300 users. Half of the users in each cell have a direct connection to the serving eNB while the rest are served by the relays (each 25\%{} of the remaining users). 
In accordance with \citep{Nokia2011}, it is assumed that each user consumes an average of 170MB of data per day in the uplink and downlink together while the uplink use being 30\%{} of the downlink one. In addition, the traffic distribution over a 24 hour period is based on the one used in \citep{Nokia2011}.
\begin{figure*}[htp]
\centering
\subfigure[Minimum residual capacity in each interval]{
\includegraphics[scale=.61]{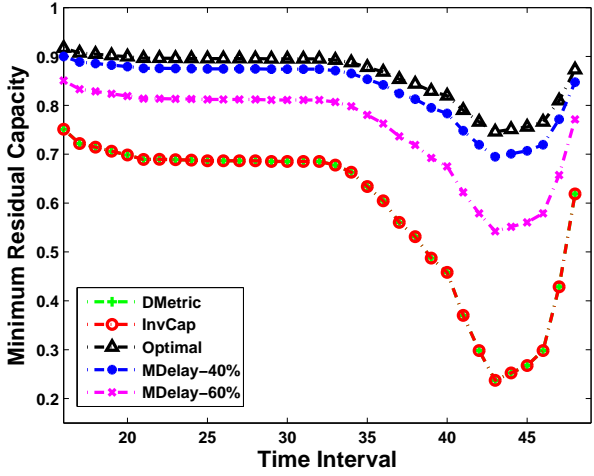}
\label{fig:subfigMaxMin}
}
\hspace*{-1.9em}
\subfigure[Minimum residual capacity ratio to optimal value ]{
\includegraphics[scale=.61]{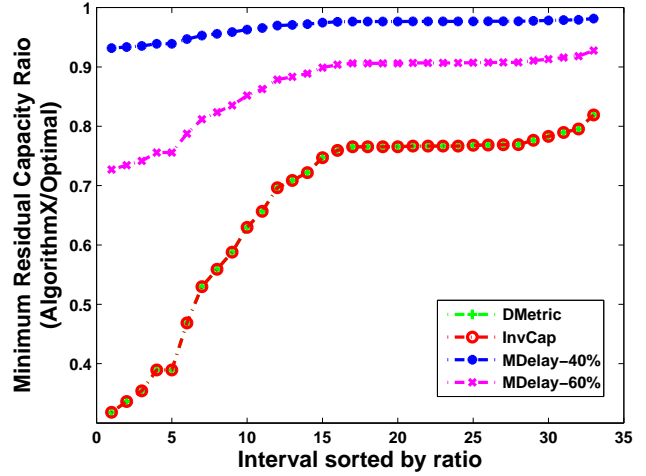}
\label{fig:subfigMaxMinR}
}
\caption[MaxMinTraffic]{Minimum residual capacity as a function of time interval }\label{fig:MaxMin}
\end{figure*}

The following metrics are used to measure the performance of the proposed scheme: (i) Minimum Residual Capacity (MRC) defined as the ratio of available bandwidth in the most congested link to the link capacity, in each time interval (ii) performance ratio defined as the ratio of each algorithm MRC to the optimal MRC, for given traffic matrices and network topology (iii) the user's perspective of the quality of the network defined as the 5-grade Mean Opinion Score (MOS) for VoIP, which in essence describes the QoE. For all the metrics, the higher values indicate more efficient resource utilization, and hence are of most \looseness=-1 interest.
The different schemes used in the performance evaluation are as follows:
\begin{itemize}
\item \textbf{DMetric}: In this algorithm, used by the Interior Gateway Protocols
(IGP), e.g., OSPF, default weights are assigned to the links in operational networks. In this work the metric assigned to eNB-eNB and relay-relay connections are set to be 10 and 100 times more than other kinds of connections (e.g., eNB-router, relay-eNB), guaranteeing the conventional operation of the LTE network. 
\item \textbf{InvCap}: In this algorithm, weights are set inversely proportional to the link capacity.
\item \textbf{MDelay}: In this algorithm, traffic demands are distributed in a given network based on the proposed scheme. Since the network topology used can guarantee the existence of minimum two disjoint paths for each node pair, we divide a traffic demand between each pair into two classes with different delay requirements: real-time and non real-time traffic. The number X next to MDelay (MDelay-X\%{}) shows the multimedia share of the total traffic between each node pair.
\item \textbf{Optimal}: In this algorithm, the optimal traffic routing for a given topology and traffic demands is calculated. The outcome is used as the baseline for our comparisons.
\end{itemize}
All the algorithms have been implemented by using the IBM ILOG CPLEX \citep{CPLEX} as the optimization solver.
%
For the first performance analysis, we make the comparison between the minimum available bandwidth (residual capacity) obtained by the proposed scheme and other routing algorithms. The higher the value, the better the achieved performance. Figure~\ref{fig:subfigMaxMin} shows the results obtained by each algorithm as a function of time. Every time interval represents a 30- minute traffic matrix, starting from 8:00AM (interval 16) to 12:00AM (interval 48). Expectedly, the optimal routing obtains the best performance, providing the maximum available resources in the bottleneck link. The worst results belong to the InvCap and DMetric, by virtue of being oblivious to the traffic type and routing all arriving flows through the shortest path between each O-D pair. Since the links belong to each type, wired and wireless, are homogeneous, having identical capacities and metrics, the results of InvCap and DMetric overlap each other. As can be seen in the figure, the proposed scheme with 40\%{} share of real-time traffic (MDelay-40\%{}) achieves close to the optimal value. In order to analyse the effect of increasing the real-time share of the total traffic, we evaluate the performance of the MDelay with 60\%{} real-time traffic between each node pair (MDelay-60\%{}). The results show that even in the case of higher share of traffic with tight-delay bounds, the proposed scheme outperforms other algorithms with a significant gap. This makes the scheme more competent for future Internet with the real-time traffic (voice and video) being deemed to be dominant. Forcing flows to chose the longest acceptable paths while keeping the shortest ones for the most eligible traffic can alleviate the effect of heterogeneous and bursty nature of multimedia traffic. 

Next, we evaluate the performance ratio of the algorithms, defined as a minimum residual capacity of each algorithm as compared to one obtained by the optimal routing (baseline). 
The time intervals are sorted in the ascending order of the performance ratio. The higher the ratio, the closer the performance of the algorithm to the optimal. The results are depicted in figure~\ref{fig:subfigMaxMinR}. While in the busy hours of a day the widely-used InvCap and DMetric can only gain 30\%{} of the optimal value, the proposed scheme can achieve a ratio more than two and three times better for 60\%{} and 40\%{} share of real-time traffic, \looseness=-1 respectively. 

The last performance metric is the user's view of the quality of the network, defined as the 5-grade MOS, for Skype calls. Recently, many studies \citep{Fiedler2010,Chen2012} have investigated the quantitative relationship between user perception of the quality of the Internet services and different network conditions. In \citep{Chen2012,Yen2013}, the MOS of the Skype call users is expressed as a logarithmic function of bitrate:
\begin{equation}
MOS = \beta +\gamma \ln(X-\alpha) \label{MOS}
\end{equation}
where $\alpha = 4.091$, $\beta =1.515$, $\gamma = 1$, and $X$ is the available bitrate in Kbps. Substituting these values, the equation returns the MOS in a single number, ranging from 1 to 5 with a higher score representing a better quality of the service. 
\begin{figure*}[tph]
\centering
\makebox[0pt][c]{%
\begin{tabular}{ccc}
\begin{minipage}{250pt}
\hspace*{-1.9em}
\frame{\includegraphics[width=258pt]{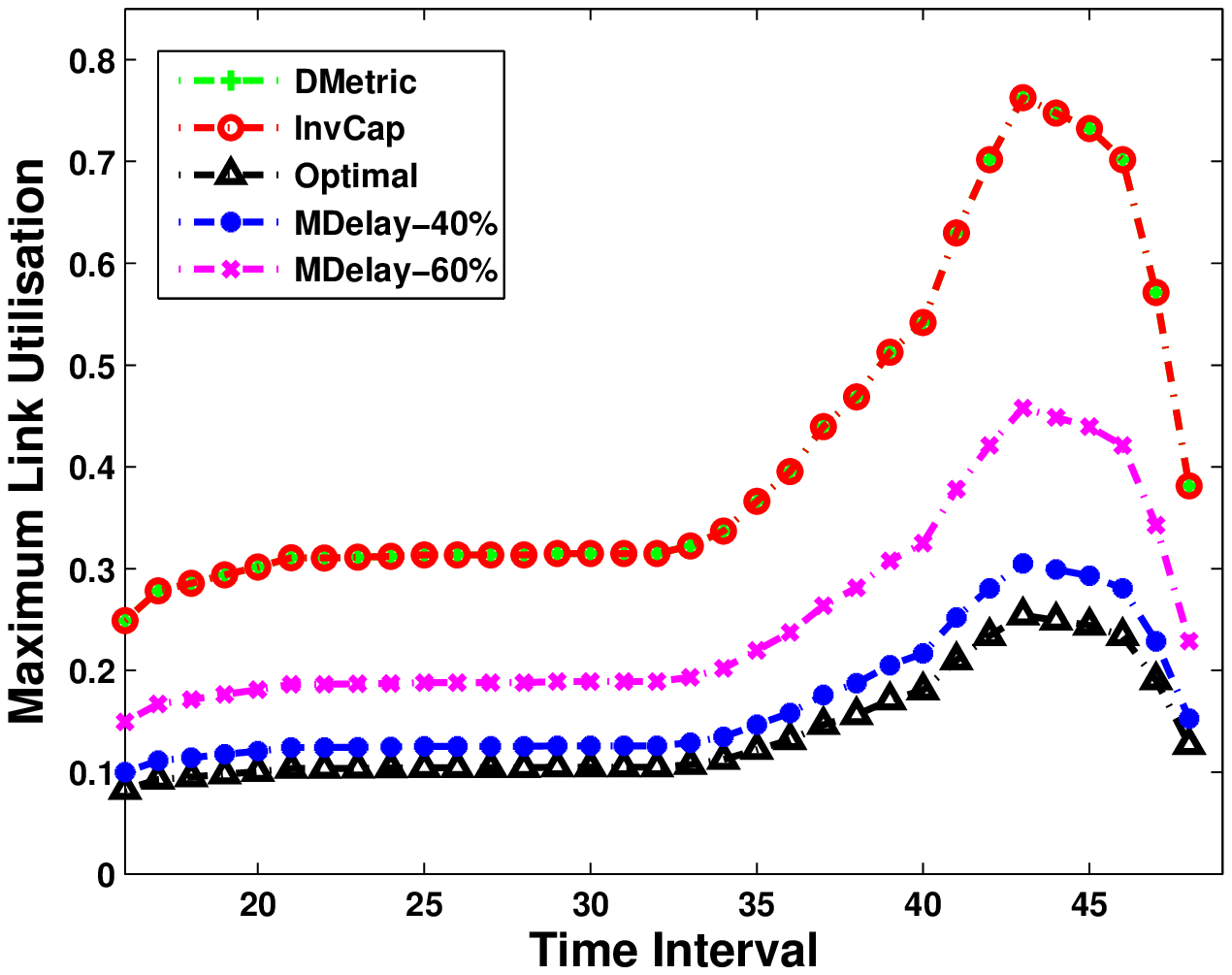}}
\caption{Maximum link utilisation}
\label{fig:MLU}
\end{minipage}
&\hspace*{-1.9em}
\begin{minipage}{250pt}
\frame{\includegraphics[width=258pt]{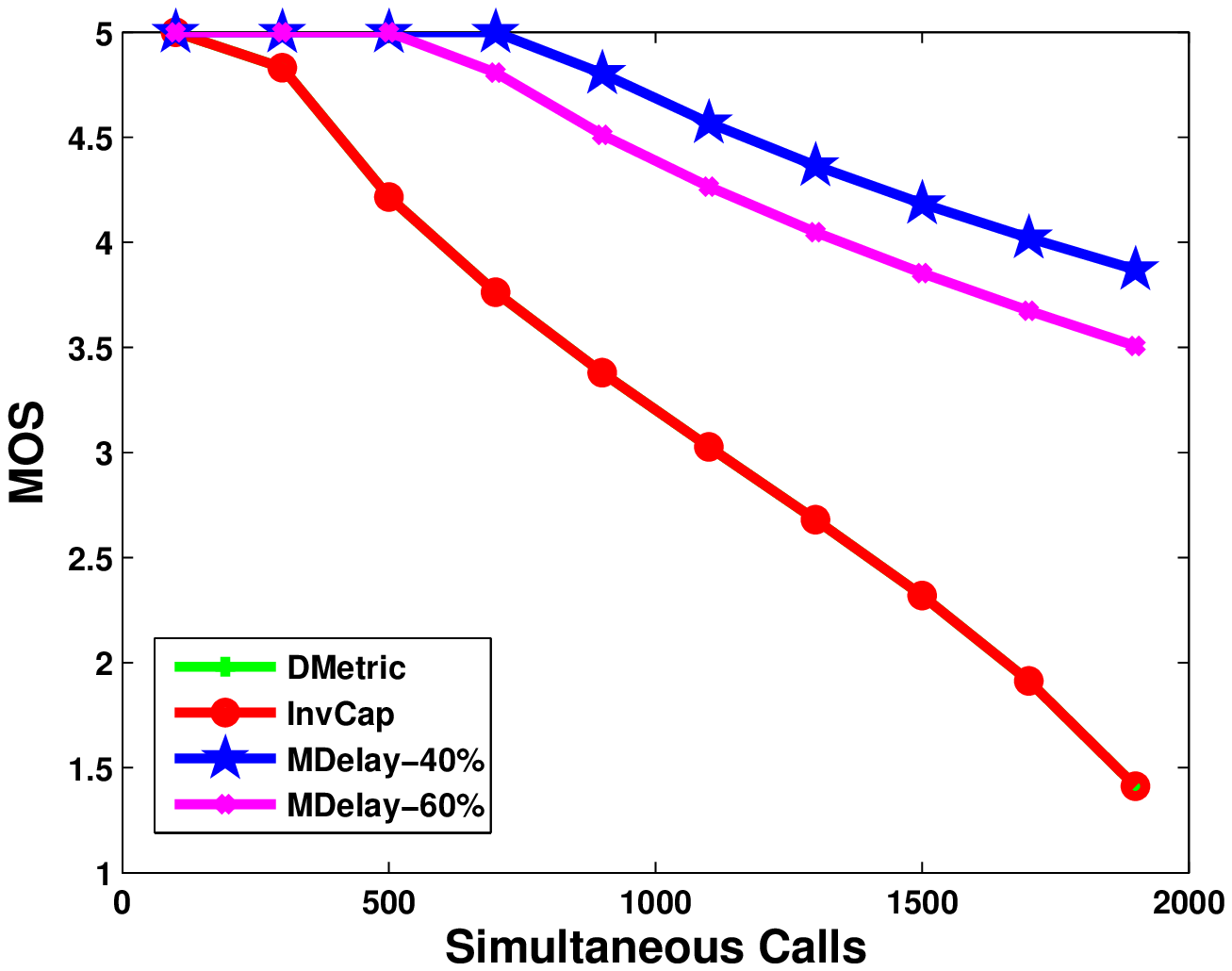}}
\caption{MOS for Skype calls}
\label{fig:QoE}
\end{minipage}
\end{tabular}
}
\end{figure*}
According to equation~(\ref{MOS}), the value of MOS is proportional to the available bandwidth between an O-D pair. Obtainable resources in each path X(P), due to the concave nature of bandwidth, can be defined as follows:
\begin{equation}
\begin{split}
\small
X(P)= Min\{ 1-MLU(s,i),..., 1-MLU(j,t)\}
\end{split}
\end{equation} 
where $MLU(i,j)$ is the maximum link utilization of link $(i,j)$. To that end, we first define the MLU of a given algorithm, in each time interval. The results obtained are shown in figure~\ref{fig:MLU}, which is expectedly the inverse of the minimum residual capacity represented in figure~\ref{fig:subfigMaxMin}. After finding $X$ for each time interval, the value is substituted in equation~(\ref{MOS-MinMax}), where the output is bounded between 1 and 5:
\begin{equation}
\begin{split}
&MOS_{min} = Max \{1,\beta +\gamma \ln(X-\alpha)\} \\
&MOS_{max} = Min \{5,\beta +\gamma \ln(X-\alpha)\}
\label{MOS-MinMax}
\end{split}
\end{equation} 
Figure~\ref{fig:QoE} shows the MOS as a function of number of simultaneous flows. As the number of users increases, due to the lower per-user bandwidth, the MOS decreases. Nevertheless, while the value of MOS for InvCap and DMetric have a sharp slope, the proposed scheme with 40\%{} real-time traffic provides a more gradual, yet still satisfactory, quality degradation. Results also show that even when the real-time share of traffic increases, the proposed scheme can provide acceptable quality for a high number of flows. 
Based on the above conducted results, the following remarks can be made: (i) the proposed scheme can provide near-optimal performance even during a peak hour traffic generation, (ii) it can achieve an acceptable performance with an increased share of real-time traffic, (iii) it takes into account the QoS requirements of the flow and (iv) it can obtain better user satisfaction, even for the high number of simultaneous calls, by providing the higher MOS. 
\section{CONCLUSIONS}
In this paper a novel QoE-centric localized mobility management scheme for future mobile networks is proposed. The scheme tries to leverage advantages of both centralized and distributed mobility management approaches, to allow enhanced performance for delay sensitive flows. To ease congestion and scalability imposed on a mobility anchor deployed within the network, optimal as well as the heuristic algorithms for path finding are presented to discover edge-disjoint logical topologies of a network, providing higher flexibility in routing. The proposed scheme allows flexibility in network setting configuration on how mobility is handled and a wide set of numerical evaluations shows that the proposed scheme achieves significant performance improvement compared to previous techniques with respect to both link utilization and user  achieved \looseness=-1 QoE.
\section*{Acknowledgements}
This work was partially supported by the FP7-CROSSFIRE Initial Training Network project funded by the European \looseness=-1 Commission.

\bibliographystyle{unsrt}
\small
\bibliography{mVoIP}
\end{document}